\documentclass[aps,prb,twocolumn,showpacs]{revtex4}
\usepackage{graphicx}
\begin{document}

\title{Infrared catastrophe in two-quasiparticle collision integral}

\author{O.\,V.\,Dimitrova$^{+*}$\/\thanks{e-mail: odimitro@ictp.it},
V.\,E.\,Kravtsov$^{+*}$} \affiliation{$^+$L.D.Landau Institute for
Theoretical Physics RAS, 117940 Moscow, Russia\\~\\ $^*$The Abdus
Salam International Centre for Theoretical Physics, P.O.B. 586,
34100 Trieste, Italy}


\date{\today}

\begin{abstract}
Relaxation of a non-equilibrium state in a disordered metal with a
spin-dependent electron energy distribution is considered. The
collision integral due to the electron-electron interaction is
computed within the approximation of a two-quasiparticle
scattering. We show that the spin-flip scattering processes with a
small energy transfer may lead to the divergence of the collision
integral for a quasi one-dimensional wire. This divergence is
present only for a spin-dependent electron energy distribution
which corresponds to the total electron spin magnetization ${\cal
M}=0$ and only for non-zero interaction in the triplet channel. In
this case a non-perturbative treatment of the electron-electron
interaction is needed to provide an effective infrared cut-off.
\end{abstract}


\maketitle

{\bf 1. Introduction.} The ground-breaking experiments by Pothier
et al. of Ref.~\onlinecite{Pothier} have demonstrated that one can
have a direct access to the non-equilibrium electron energy
distribution function $f(E)$ and through it to the inelastic
collision integral ${\cal K}_{\rm coll}(E)$ which enters the
kinetic equation~\cite{foot1}:
\begin{equation}
\label{KE}
\partial_{t}f(E;x,t)-D\nabla^{2}f(E;x,t)=-{\cal K}_{\rm
coll}(E;x,t).
\end{equation}
In turn, studying the collision integral
gives one an important information on interaction and dynamics of
quasiparticles in a dirty metal. In this way the predictions of
the theory of electron interaction in disordered
metals~\cite{AA-book,AAKh} were checked~\cite{Pothier,PBir} and an
unexpected strong sensitivity of the energy relaxation to the
presence of magnetic impurities~\cite{GlazKam} was established.

\begin{figure}
\includegraphics[angle=0,width=3.1in]{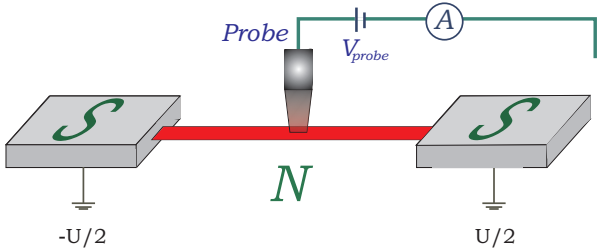}
\caption{\label{Set} The setup: a quasi-one dimensional disordered
normal-metal wire connected through the weak tunnel junctions to
the two superconducting leads. The non-equilibrium is created by
applying the finite bias voltage $U$ between the leads.}
\end{figure}

The main idea of Ref.~\onlinecite{Pothier} was to use the sharp
features in the energy dependence of the density of states (DoS)
$\nu_{\rm probe}(E)$ of a superconducting probe electrode, which
enabled to extract $f(E)$ by measuring the differential
conductance of the tunnel junction between the normal metal sample
and the probe electrode. Recently, the same idea has been
suggested~\cite{Gia} to create a non-equilibrium {\it
spin-dependent} electron energy distribution $f_{\sigma}(E)$ and
thereby to obtain a spin-polarized current through the probe. The
sketch of the experimental setup is shown in Fig.~\ref{Set}. The
sample in a form of a quasi-one dimensional disordered
normal-metal wire is connected through the weak tunnel junctions
to the two superconducting leads. The non-equilibrium is created
by applying the finite bias voltage $U$ between the leads. The
spin dependence of $f_{\sigma}(E)$ ($\sigma=\pm 1$ for the spin
projections $\uparrow (\downarrow)$) is caused by magnetic field
applied to the superconducting leads with the DoS
$\nu^{(R)}_{\sigma}(E)=\nu_{S}(E+U/2\pm\sigma E_{Z})$ and
$\nu^{(L)}_{\sigma}(E)=\nu_{S}(E-U/2+\sigma E_{Z})$ where
$\nu_{S}(E)=\nu\, \Re\left[ E/\sqrt{E^{2}-\Delta^{2}}\right] $ and
the Zeeman shift $E_{Z}=\mu_{B}H$ is taken with the sign $\pm$
depending on whether the directions of the magnetic field in the
right and left leads are parallel or anti-parallel. In the absence
of relaxation $f_{\sigma}(E)$ is given by~\cite{Gia,Klap}:
\begin{equation}
\label{f-sig}
f_{\sigma}(E)=\frac{\nu^{(L)}_{\sigma}(E)f_{F}(E-U/2)+\nu^{(R)}_{\sigma}(E)f_{F}(E+U/2)}{\nu^{(L)}_
{\sigma}(E)+\nu^{(R)}_{\sigma}(E)},
\end{equation}
where  $f_{F}(E)$ is the Fermi distribution function. The measured
quantity is the differential conductance with respect to the probe
bias $V_{\rm probe}$ across the probe tunnel contact. The probe
contact can act as a spin-analyzer provided an additional magnetic
field is also applied to a superconducting probe electrode.

\begin{figure}
\includegraphics[angle=0,width=3.1in]{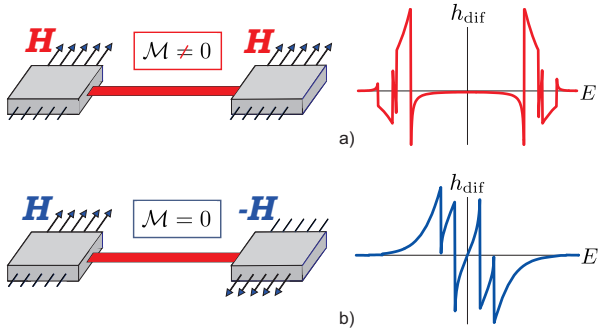}
\caption{\label{Mdiff} Two distinct cases: (a) with parallel and
(b) with anti-parallel magnetic fields in the superconducting
leads. In (a) case a non-equilibrium state with a nonzero total
spin polarization is created, while in (b) case ${\cal M}=0$. The
typical form of the difference $h_{\rm
dif}=h_{\uparrow}(E)-h_{\downarrow}(E)$ is shown in both cases. }
\end{figure}

There are two distinct cases schematically shown in
Fig.~\ref{Mdiff}a and Fig.~\ref{Mdiff}b: (i) with parallel and
(ii) with anti-parallel magnetic fields in the superconducting
leads. In the former case a non-equilibrium state with a nonzero
total spin polarization
\begin{equation}
\label{S} {\cal M}=\int dE\, [f_{\uparrow}(E)-f_{\downarrow}(E)]
\end{equation}
is created, while in the latter case ${\cal M}=0$. The typical
form of the difference $f_{\rm dif
}=f_{\uparrow}(E)-f_{\downarrow}(E)$ that follows from
Eq.~(\ref{f-sig}) is shown in Fig.~\ref{Mdiff} in both cases.

The aim of this work is to consider the relaxation of such a
spin-dependent distribution caused by the electron-electron
interaction. For this we derive the collision integral ${\cal
K}_{\rm coll }$ in the approximation of the two-quasiparticle
collisions in the case where both the spin-singlet and the
spin-triplet channel of the interaction are present. The detailed
derivation of the collision integral due to the electron-electron
interaction has been recently carried out in
Refs.~\onlinecite{AhCatAl, CatAl}. However, the analysis has been
limited to the case of the spin-independent distribution functions
$f(E)$, while we are going to focus on the relaxation of the
difference $f_{\rm dif}(E)$. The results of calculation of the
collision integral for the spin-dependent distribution function
have been also recently reported in Ref.~\onlinecite{Schel-Bur}.
However, the authors considered a limited class of distributions
with a very particular spin-dependence equivalent to a shift in
the energy $f_{\uparrow}(E)=f_{\downarrow}(E+\delta E)$. Such type
of dependence does not hold e.g. in the experimental setup of
Figs.~\ref{Set},~\ref{Mdiff}.

The main qualitative result of our analysis is that there are
three different contributions in the collision integral. Two of
them are also present if only the singlet channel of the
interaction is considered, with only their amplitudes depending on
the triplet channel interaction constant $F$. The third
contribution corresponding to the spin-flip process is only
present when the triplet channel of the interaction is taken into
account. Its magnitude depends~\cite{Schel-Bur} essentially on the
conserving total spin ${\cal M}$. However more importantly, it is
singular for the non-equilibrium spin-dependent distribution with
${\cal M}=0$ which naturally arises in the experimental situation
(ii) of Fig.~\ref{Mdiff}b. The existence of such a singularity
which never occurs  if only the singlet channel of the interaction
is present, is the main qualitative result of this work.

\begin{figure}
\includegraphics[angle=0,width=3.1in]{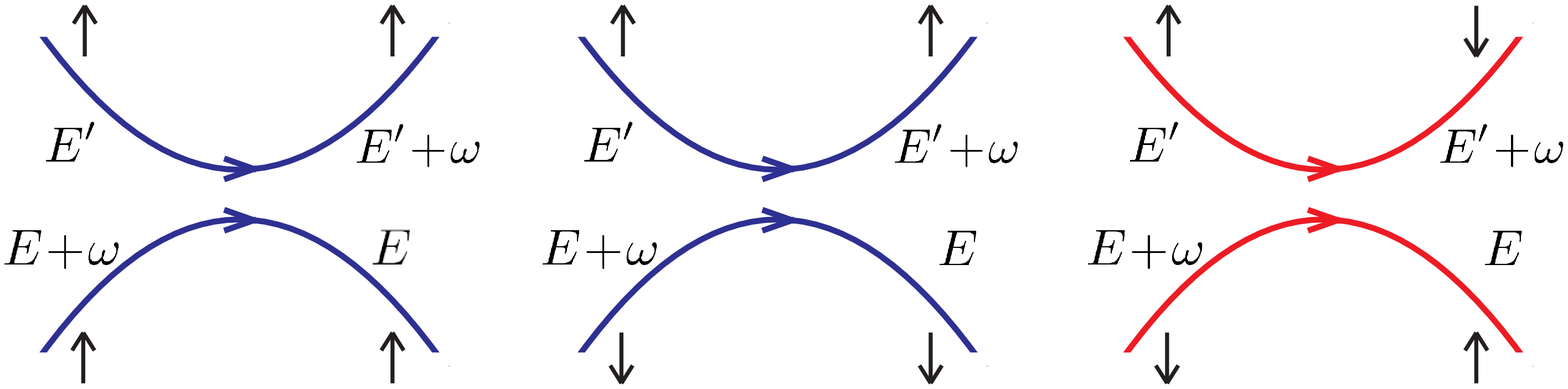}
\caption{\label{channels} Three possible processes, allowed by
conservation laws. In the initial state the quasiparticles have:
(i) the same spin projection which remain unchanged during the
collision, (ii) opposite spin projections which do not change
during the collision, (iii) opposite spin projections and the
collision results in a spin-flip of both quasiparticles.}
\end{figure}

{\bf 2.Three contributions to the collision integral.} For a
generic two-quasiparticle collision in a disordered metal in the
absence of spin-orbit interaction and magnetic impurities two
quantities are conserved: the total energy ${\cal E}$ and the
total spin  ${\cal M}$. The latter conservation law allows only
three possible processes (see Fig.~\ref{channels}): (i) in the
initial state the quasiparticles have the same spin projections
$(1/2)\sigma$ which remain unchanged during the collision, (ii) in
the initial state the quasiparticles have opposite spin
projections which do not change during the collision, (iii) in the
initial state the quasiparticles have opposite spin projections
and the collision results in a spin-flip of both quasiparticles.
Each process corresponds to a certain term in the collision
integral that contains combinations of the type $f^{\rm
in}(E+\omega)\,f^{\rm in'}(E')\,[1-f^{\rm fin}(E)]\,[1-f^{\rm
fin'}(E'+\omega)]-[1-f^{\rm in}(E+\omega)]\,[1-f^{\rm
in'}(E')]\,f^{\rm fin}(E)\,f^{\rm fin'}(E'+\omega)$ which for the
processes (i)-(iii) take, respectively, the forms:
\begin{eqnarray}
\label{comb1} I^{(1)}_{\sigma}&=&\int
dE'\,\left[-(1-h_{\sigma,E}h_{\sigma,E+\omega})(h_{\sigma,E'+\omega}-h_{\sigma,E'})\right.
\nonumber \\ &+& \left
(h_{\sigma,E+\omega}-h_{\sigma,E})(1-h_{\sigma,E'}h_{\sigma,E'+\omega})
\right],\\ \label{comb2}I^{(2)}_{\sigma}&=&\int
dE'\,\left[-(1-h_{\sigma,E}h_{\sigma,E+\omega})(h_{-\sigma,E'+\omega}-h_{-\sigma,E'})\right.
\nonumber \\ &+& \left
(h_{\sigma,E+\omega}-h_{\sigma,E})(1-h_{-\sigma,E'}h_{-\sigma,E'+\omega})
\right],\\
\label{comb3}I^{(3)}_{\sigma}&=& \int
dE'\,\left[-(1-h_{\sigma,E}h_{-\sigma,E+\omega})(h_{-\sigma,E'+\omega}-h_{\sigma,E'})\right.
\nonumber \\&+& \left
(h_{-\sigma,E+\omega}-h_{\sigma,E})(1-h_{\sigma,E'}h_{-\sigma,E'+\omega})
\right],
\end{eqnarray}
where $h_{\sigma,E}=1-2f_{\sigma}(E)$.

The collision integral ${\cal K}_{\rm coll}(\sigma,E)$ can be
represented as follows:
\begin{equation}
\label{coll-I} {\cal K}_{\rm coll}(\sigma,E)=\sum_{p=1}^{3}\int
\frac{d\omega}{2\pi\nu}\,K_{p}^{\sigma}(\omega)I^{(p)}_{\sigma}(E,\omega),
\end{equation}
where $\nu$ is the DOS (per spin direction) of the normal-metal
sample at the Fermi level. The quantities $K_{p}^{\sigma}(\omega)$
describe the strength of relaxation due to the corresponding
processes (the quantities corresponding to the singlet channel
$K_{1,2}^\sigma (\omega)=K_{1,2} (\omega)$ does not depend on
${\cal M}$ and hence on $\sigma$). We obtained the following
expressions for them valid in the limit $p_{F}\ell\gg 1$ ($p_{F}$
is the Fermi momentum, $\ell$ is the elastic scattering length)
and for the diffusive quasiparticle dynamics:
\begin{eqnarray}
\label{K1} K_{1}(\omega)&=&\frac{1}{{\cal V}}\sum_{{\bf
q}}\frac{1}{\omega^{2}+(D{\bf q
}^{2})^{2}}\,\frac{\left(\frac{1}{2}+F\right)^{2}+\frac{\omega^{2}}{(2D{\bf
q }^{2})^{2}}}{(1+F)^{2}+\frac{\omega^{2}}{(D{\bf q}^{2})^{2}}},\\
\label{K2}K_{2}(\omega)&=&\frac{1}{{\cal V}}\sum_{{\bf
q}}\frac{1}{\omega^{2}+(D{\bf q
}^{2})^{2}}\,\frac{\frac{1}{4}+\frac{\omega^{2}}{(2D{\bf q
}^{2})^{2}}}{(1+F)^{2}+\frac{\omega^{2}}{(D{\bf q}^{2})^{2}}},\\
\label{K3}K_{3}^{\sigma}(\omega)&=&\frac{1}{{\cal V}}\sum_{{\bf
q}}\frac{1}{\omega^{2}+(D{\bf q
}^{2})^{2}}\,\frac{F^{2}}{(1+F)^{2}+\frac{(\omega-\sigma F{\cal
M})^{2}}{(D{\bf q}^{2})^{2}}}.\nonumber \\
\end{eqnarray}
In Eqs.~(\ref{K1})-(\ref{K3}) by $F$ ($F=-1$ corresponds to the
Stoner instability) we denoted the Fermi-liquid constant
corresponding to the triplet channel of the electron-electron
interaction. The summation over ${\bf q}$ can be replaced by
integration $\frac{1}{{\cal V}}\sum_{q}\rightarrow \int
\frac{d^{d}q}{(2\pi)^{d}}$ in the limit $\omega\gg E_{Th}\equiv
D/L^{2}$ ($D$ is the diffusion coefficient, $L$ is the length of
the disordered sample) which will be considered below.

Note that at small $F\ll1$ we obtain up to the linear in $F$
terms: $K_{3}=0$ and
\begin{equation}
\label{small} K_{1,2}=\frac{1}{4{\cal
V}}\sum_{q}\frac{1}{\omega^{2}+(D{\bf q }^{2})^{2}}\,\left[1\pm
\frac{2F}{1+\frac{\omega^{2}}{(D{\bf q}^{2})^{2}}} \right].
\end{equation}
In this limit the spin-spin interaction in the triplet channel
results in only a small (and opposite in sign for the parallel and
anti-parallel spins of the two quasiparticles in the initial
state) change in the amplitudes of the processes (i) and (ii)
which are dominated by the singlet channel of the
electron-electron interaction. Note that under the restrictions on
the form of the spin-dependence of $h_{\sigma,E}$ adopted in
Ref.~\onlinecite{Schel-Bur} the combinations $I^{(1)}_{\sigma}$
and $I^{(2)}_{\sigma}$ appeared to be identical. This is why the
result of Ref.~\onlinecite{Schel-Bur} contained only the
combination $(K_{1}+K_{2})I^{(1)}_{\sigma}$ and $K_{3}^\sigma
I^{(3)}_{\sigma}$.

{\bf 3. Relaxation of a non-equilibrium distribution and the
conservation laws.} As has been already mentioned, the form of the
collision integral should be compatible with the two conservation
laws. The conservation of the total energy requires:
\begin{equation}
\label{consEn} \int dE\; E\,[{\cal K}_{\rm coll}(\uparrow,E)+{\cal
K}_{\rm coll}(\downarrow, E)]=0.
\end{equation}
The conservation of the total spin polarization leads to:
\begin{equation}
\label{consSp} \int dE\; [{\cal K}_{\rm coll}(\uparrow, E)- {\cal
K}_{\rm coll}(\downarrow, E)]=0.
\end{equation}
One can check using Eqs.~(\ref{comb1})-(\ref{K3}) that fulfillment
of those two conservation laws is guaranteed by the structure of
$I^{(p)}_{\sigma}(E,\omega)$ and the following properties of the
kernels $K_{p}(\omega)$:
\begin{equation}
\label{conserv}
K_{1,2}(\omega)=K_{1,2}(-\omega),\;\;\;\;K_{3}^\sigma(\omega)=
K_{3}^{-\sigma}(-\omega).
\end{equation}
What we would like to note here is that the full relaxation to
equilibrium due to the electron-electron interaction is only
possible if ${\cal M}=0$. Indeed, the fixed solutions to the
kinetic equation Eq.~(\ref{KE}) which correspond to all
combinations Eq.~(\ref{comb1})-(\ref{comb3}) vanishing
identically, are the Fermi distribution functions
$f_{\uparrow(\downarrow)}^{(0)}(E)=f_{F}(E\mp{\cal M}/2)$. Any
non-equilibrium distribution tends to relax to these fixed
solutions. However only at ${\cal M}=0$ we have
$f_{\uparrow}^{(0)}(E)\equiv f_{\downarrow}^{(0)}(E)$ which
corresponds to the complete equilibrium. So we encounter for the
first time with the special role of the ${\cal M}=0$ condition.

 {\bf 4. Collision
integral for a quasi-1d wire and the infrared catastrophe at
${\cal M}=0$.} For the quasi-1d experimental geometry of
Fig.~\ref{Set}, Eqs.~(\ref{K1})-(\ref{K3}) can be
straightforwardly evaluated:
\begin{eqnarray}
\label{K1-2-one} K_{1,2}(\omega)=
\frac{C_{1,2}}{8\,S\,\sqrt{2D\,(1+F)}}\,\frac{1}{|\omega|^{3/2}},
\end{eqnarray}
\begin{eqnarray}
\label{K3-one}K_{3}^{\sigma}(\omega)&=&
\frac{C_{3}}{S\,(\sqrt{|\omega-\sigma F{\cal M
}|}+\sqrt{|\omega|}\sqrt{1+F})} \\
&\times&\frac{1}{[|\omega-\sigma F{\cal
M}|+|\omega|(1+F)]\,\sqrt{2D(1+F)}},\nonumber
\end{eqnarray}
where $S$ is the cross-section area of the quasi-1d wire,
\begin{equation}
\label{C1} C_{1}=\left(1+\frac{4F(1+F)}{(1+\sqrt{1+F})(2+F)}
\right),
\end{equation}
$C_{2}=1$, $C_{3}=F^{2}/2$ and the legitimate values of $F$ are
$F>-1$.

A remarkable property~\cite{Schel-Bur} of $K_{3}^\sigma$ is that
it depends on the spin polarization ${\cal M}$. For ${\cal M}\neq
0$ $K_{3}^\sigma(\omega)$ is finite at all values of $\omega$ and
the collision integral is well defined. A peculiar situation
arises when ${\cal M}=0$. In this case
$K_{3}^\sigma(\omega)\propto \frac{1}{|\omega|^{3/2}}$ is singular
at $\omega=0$. At the first glance there is nothing special in
such a degeneracy which is the same as for the standard case of
the electron-electron interaction in the singlet
channel~\cite{Pothier, AA-book}. The difference, however, is in
the form of $I^{(3)}_{\sigma}$ as compared to
$I^{(1,2)}_{\sigma}$. As is seen from
Eqs.~(\ref{comb1})-(\ref{comb3}), the cancellations of the ``in''
and ``out'' terms lead to: $I^{(1,2)}_{\sigma}(\omega=0)=0$ for
any $f_{\sigma}(E)$, while $I^{(3)}_{\sigma}(\omega=0)\neq 0$
unless $f_{\rm dif }=f_{\uparrow}(E)-f_{\downarrow}(E)$ is
identically zero. This means that the singularity
$K_{p}(\omega)\propto\frac{1}{|\omega|^{3/2}}$ is not dangerous
for terms proportional to $I^{(1,2)}_{\sigma}$ (which are the only
terms that arise in the case of electron-electron interaction in
the singlet channel) but it leads to the divergency of the term
$\propto I^{(3)}_{\sigma}$ corresponding to the spin-flip
processes due to the electron-electron interaction in the triplet
channel. So we have an infrared catastrophe in the collision
integral in the case of a spin-dependent electron energy
distribution with ${\cal M }=0$.

{\bf 5. Derivation of the collision integral.} Before we discuss
the physical origin of such a catastrophe and the ways to cure it
we briefly outline the derivation of the collision integral
Eqs.~(\ref{comb1})-(\ref{comb3}),(\ref{K1})-(\ref{K3}).

Following the original work of Keldysh~\cite{Keldysh} we represent
the collision integral as follows:
\begin{equation}
\label{keld} {\cal K}_{\rm coll}=-i(\check{\Sigma}\check{{\cal
G}}-\check{{\cal G}}\check{\Sigma})_{12}
\end{equation}
where
\begin{eqnarray}\label{matrix}
\check\Sigma= \left( \begin{array}{cc} \displaystyle \Sigma^R&
\Sigma^K
\\ \displaystyle
0& \Sigma^A
\end{array} \right),\qquad
\check{\cal G}= \left( \begin{array}{cc} \displaystyle {\cal G}^R&
{\cal G}^K
\\ \displaystyle
0& {\cal G}^A
\end{array} \right)
\end{eqnarray}
and $\Sigma^{R,A,K}$ and ${\cal G}^{R,A,K}$ are retarded, advanced
and Keldysh components of the self-energy part and an exact
single-particle Green's function. In the two-quasiparticle
collision approximation we adopt in this paper, the self-energy
part is given by:
\begin{eqnarray}
\label{sigma} \Sigma^{R}&=&{\cal
D}^{K}_{\omega}\,G^{R}_{E+\omega}+{\cal
D}^{A}_{\omega}\,G^{K}_{E+\omega},\\
\Sigma^{A}&=&{\cal D}^{K}_{\omega}\,G^{A}_{E+\omega}+{\cal
D}^{R}_{\omega}\,G^{K}_{E+\omega},\\
\Sigma^{K}&=&{\cal D}^{K}_{\omega}\,G^{K}_{E+\omega}+({\cal
D}^{A}_{\omega}-{\cal
D}^{R}_{\omega})\,(G^{R}_{E+\omega}-G^{A}_{E+\omega}),
\end{eqnarray}
where the integration over $\omega$ and a proper summation over
spin indices $\alpha,\beta,\gamma,\delta$ is assumed in all three
equations.

In this approximation we neglect (a) the interaction corrections
to the vertex part $\Gamma=1$ and (b) the interaction corrections
to the single-particle Green's function ${\cal G}_{E+\omega}$
which is supposed to be equal to the corresponding Green's
function without interaction $G_{E+\omega}$. In Eqs.~(\ref{sigma})
we denote by ${\cal D}_{\omega}$ the {\it dynamically screened}
interaction:
\begin{equation}
\label{calD} {\cal
D}_{\alpha\beta,\gamma\delta}=\hat{\sigma}^{i}_{\alpha\beta}D_{ij}\hat{\sigma}^{j}_{\gamma\delta},
\end{equation}
where $\hat{\sigma}^{i}$ ($i=1,2,3$) are the Pauli matrices
$\hat{\sigma}_{x,y,z}$ and $\hat{\sigma}^{4}$ is the unit matrix.

The $4\times 4$ interaction matrix $D^{R(A)}_{ij}$ obeys an
RPA-like equation:
\begin{equation}
\label{RPA-RA}
D^{R(A)}_{ij}=U_{i}\delta_{ij}+U_{i}\delta_{ik}\Pi^{R(A)}_{kl}D^{R(A)}_{lj},
\end{equation}
where $U_{i}$ ($i=1,2,3$) is the bare interaction constant
$F^{\sigma}=F/(4\nu)$ of the electron-electron interaction in the
triplet channel:
\begin{eqnarray}
\label{trip-int}
\hat{H}_{\sigma}&=&\frac{1}{2}\sum_{p_{1,2},q<q^{*}}\sum_{j=1,2,3}
\sum_{\alpha\beta\gamma\delta}F^{\sigma}\\
\nonumber &\times&
[\psi^{\dagger}_{\alpha}(\vec{p}_{1})\,\hat{\sigma}^{j}_{\alpha\beta}\,\psi_{\beta}(\vec{p}_{1}-\vec{q})]\\
\nonumber &\times&
[\psi^{\dagger}_{\gamma}(\vec{p}_{2})\,\hat{\sigma}^{j}_{\gamma\delta}\,\psi_{\delta}(\vec{p}_{2}+\vec{q})]
\end{eqnarray}
and $U_{4}\equiv U^{\rho}=F^{\rho}+V(q)$ is the bare interaction
in the singlet channel which at small $q$ is dominated by the
Coulomb interaction $V(q)\rightarrow\infty$:
\begin{eqnarray}
\label{singl-int}
\hat{H}_{\rho}&=&\frac{1}{2}\sum_{p_{1,2},q<q^{*}}
\sum_{\alpha\gamma}
[V(q)+F^{\rho}]\\
\nonumber &\times&
[\psi^{\dagger}_{\alpha}(\vec{p}_{1})\,\psi_{\alpha}(\vec{p}_{1}-\vec{q})]
\;[\psi^{\dagger}_{\gamma}(\vec{p}_{2})\,\psi_{\gamma}(\vec{p}_{2}+\vec{q})].
\end{eqnarray}
The generalized polarization bubbles are given by
\begin{equation}
\label{Pi}
\Pi^{R,A,K}_{kl}=\sum_{\alpha,\beta}\hat{\sigma}^{k}_{\alpha\beta}\pi^{R,A,K}_{\beta\alpha}\hat{\sigma}^{l}_{\beta\alpha},
\end{equation}
where the retarded and advanced polarization bubbles
$\pi^{R(A)}_{\alpha\beta}\rightarrow
G^{R(A)}_{\alpha\alpha}G^{K}_{\beta\beta}+G^{K}_{\alpha\alpha}G^{A(R)}_{\beta\beta}$
for the spin-independent single particle Hamiltonian (no magnetic
impurities and no spin-orbit interaction) contain retarded,
advanced and Keldysh Green's functions.

The Keldysh component $D^{K}_{ij}$ of the dynamically screened
interaction is expressed explicitly through $D^{R}_{ij}$ and
$D^{A}_{ij}$:
\begin{equation}
\label{D-K} D^{K}_{ij}=D^{R}_{ik}\Pi^{K}_{kl}D^{A}_{lj},
\end{equation}
where $\Pi^{K}_{ij}$ is given by Eq.~(\ref{Pi}) but with
$\pi^{K}_{\alpha\beta}\rightarrow
G^{K}_{\alpha\alpha}G^{K}_{\beta\beta}+
G^{R}_{\alpha\alpha}G^{A}_{\beta\beta}+G^{A}_{\alpha\alpha}G^{R}_{\beta\beta}$.

The central point of the derivation of the collision integral is
the ansatz that involves the non-equilibrium electron energy
distribution function $h_{\sigma,E}$:
\begin{equation}
\label{anz}
G^{K}_{E,\sigma}=(G^{R}_{E,\sigma}-G^{A}_{E,\sigma})\,h_{\sigma,E}.
\end{equation}
The similar ansatz has been suggested by Keldysh~\cite{Keldysh}
for the {\it exact} Green's functions. We will be using
Eq.~(\ref{anz})
 for the Green's functions {\it without
electron-electron interaction}. The reason is that the
perturbation theory in interaction can be built using
Eq.~(\ref{anz}) with an arbitrary ``initial'' distribution
function $h^{(0)}_{\sigma,E}$ compatible with the Fermi
statistics. It will cancel out anyway in the final result as the
initial distribution has to be forgotten in the non-equilibrium
steady state. Diagrammatically this cancellation happens because
of the proliferation of singular ``loose diffusons''~\cite{KY}
which, however, make impossible the perturbative analysis. There
is only one single choice of $h^{(0)}_{\sigma,E}$ in
Eq.~(\ref{anz}) -- the true solution $h_{\sigma,E}$ of the kinetic
equation -- when such proliferation does not occur and all the
diagrams with loose diffusons are equal to zero~\cite{KY}.

Using Eq.~(\ref{anz}) one can obtain the following expressions for
the polarization bubbles:
\begin{eqnarray}\label{bubbles}
\pi^{R}_{\alpha\beta}&=&-i\,\left[(h_{\alpha,E'+\omega}-h_{\beta,E'})G^{R}_{E'+\omega}G^{A}_{E'}
-2i\nu\right], \\
\pi^{A}_{\alpha\beta}&=&+i\,\left[(h_{\alpha,E'+\omega}-h_{\beta,E'})G^{A}_{E'+\omega}G^{R}_{E'}
+2i\nu\right], \\
\pi^{K}_{\alpha\beta}&=&i\,(1-h_{\alpha,E'+\omega}h_{\beta,E'})\Delta
G_{E'+\omega}\Delta G_{E'},
\end{eqnarray}
where $\Delta G_{E}=(G^{R}-G^{A})_{E}$ and an integration over
$\int dE'/2\pi$ is assumed. The next step is the standard disorder
average of the product $G^{R}_{E'+\omega}G^{A}_{E'}$ with the
result
\begin{equation}
\label{av} \langle G^{R}_{E'+\omega}G^{A}_{E'}\rangle_{\omega,{\bf
q}}=2\pi\nu/(D{\bf q}^{2}-i\omega).
\end{equation}
This eliminates the dependence on $E'$ everywhere but in the
distribution functions $h_{\alpha,E'}$.

At this point it is appropriate to note on the difference between
$\pi^{R(A)}_{\alpha\alpha}$ (or the spin-independent case) and
$\pi^{R(A)}_{\alpha\beta}$ with $\alpha\neq\beta$. In the former
case one can use an identity
\begin{equation}
\label{iden} \int
dE'\,(h_{\alpha,E'+\omega}-h_{\alpha,E'})=2\omega,
\end{equation}
which holds for an arbitrary function $h_{\alpha,E'}$ which at
$|E'|\rightarrow\infty$ converges sufficiently fast to
$\mathrm{sign}(E')$. Then one immediately obtains the standard
result which is independent of the electron energy distribution
function:
\begin{equation}
\label{pira} \pi^{R}_{\alpha
\alpha}\equiv\pi^{R}=-2\nu\,\frac{D{\bf q}^{2}}{D{\bf
q}^{2}-i\omega},\;\;\;\;\pi^{A}_{\alpha\alpha}=(\pi^{R})^{*}.
\end{equation}
In contrast to that, the corresponding integral in
$\pi^{R}_{\alpha,-\alpha}$ contains a part $\int
dE'\,(h_{\alpha,E'}-h_{-\alpha,E'})$ proportional to the total
spin polarization ${\cal M}$ of the non-equilibrium state given by
Eq.~(\ref{S}). Thus we obtain:
\begin{equation}
\label{piRA}
\pi^{R}_{\alpha,-\alpha}=\pi^{A*}_{\alpha,-\alpha}=\pi^{R}\pm
2i\frac{\nu{\cal M}}{D{\bf q}^{2}-i\omega},
\end{equation}
where the sign $\mp$ corresponds to $\alpha=\uparrow(\downarrow)$.
For completeness we also give an expression for
$\pi^{K}_{\alpha\beta}$:
\begin{equation}
\label{pika} \pi^{K}_{\alpha\beta}=-2i\nu\,\frac{D{\bf
q}^{2}}{(D{\bf q}^{2})^{2}+\omega^{2}}\;\int
dE'\,(1-h_{\alpha,E'+\omega}h_{\beta,E'}).
\end{equation}
The dependence of Eqs.~(\ref{piRA}),(\ref{pika}) on the spin
projections $\alpha,\beta$ makes the $4\times4$ matrices
$\Pi^{R,A,K}_{kl}$ non-diagonal:
\begin{eqnarray}\label{matrPi}
\Pi^{R,A,K}_{ik}=\left( \begin{array}{cccc} \displaystyle \Pi_2 &
-i \Pi_3 & 0 & 0
\\ \displaystyle
i\Pi_3 & \Pi_2& 0 & 0
\\ \displaystyle
0 & 0 & \Pi_0 & \Pi_1\\ \displaystyle 0 & 0 & \Pi_1 & \Pi_0
\end{array} \right)^{R,A,K},
\end{eqnarray}
where for any of the omitted superscripts $R,A,K$
\begin{eqnarray}
\label{PiPi}
\Pi_{0}&=&\pi_{\uparrow\uparrow}+\pi_{\downarrow\downarrow},\;\;\;\;\Pi_{1}=\pi_{\uparrow\uparrow}-\pi_{\downarrow\downarrow},\\
\nonumber
\Pi_{2}&=&\pi_{\uparrow\downarrow}+\pi_{\downarrow\uparrow},\;\;\;\;\Pi_{3}=\pi_{\uparrow\downarrow}-\pi_{\downarrow\uparrow}.
\end{eqnarray}
Correspondingly, the $4\times4$ matrices $D^{R,A,K}_{ij}$ found
from Eqs.~(\ref{RPA-RA}),(\ref{D-K}) appear to have the
off-diagonal structure similar to Eq.~(\ref{matrPi}). For
$D^{R(A)}=(1-U\Pi^{R(A)})^{-1}U$ we obtain:
\begin{eqnarray}\label{matrD}
D^{R,A,K}_{ik}=\left( \begin{array}{cccc} \displaystyle D_2 & -i
D_3 & 0 & 0
\\ \displaystyle
i D_3 & D_2& 0 & 0
\\ \displaystyle
0 & 0 & D_{zz} & D_{z0}\\ \displaystyle 0 & 0 & D_{0z} & D_{00}
\end{array} \right)^{R,A,K},
\end{eqnarray}
with
\begin{eqnarray}
\label{DD-RA}
D_{2}^{R(A)}&=&F^{\sigma}\,(1-F^{\sigma}\Pi_{2}^{R(A)})\;\zeta^{R(A)},\\
\nonumber
D_{3}^{R(A)}&=&(F^{\sigma})^{2}\,\Pi_{3}^{R(A)}\,\zeta^{R(A)},\\
\nonumber
\zeta^{R(A)}&=&\frac{1}{(1-2F^{\sigma}\pi_{\uparrow\downarrow}^{R(A)})(1-2F^{\sigma}\pi^{R(A)}_{\downarrow\uparrow})},
\end{eqnarray}
\begin{eqnarray}
\label{DD-RA1}
D_{zz}^{R(A)}&=&\frac{F^{\sigma}}{1-2F^{\sigma}\pi^{R(A)}},\\
\nonumber
D_{00}^{R(A)}&=&\frac{U^{\rho}}{1-2U^{\rho}\pi^{R(A)}}\rightarrow-\frac{1}{2\pi^{R(A)}},\\
\nonumber D^{R(A)}_{z0}&=&D^{R(A)}_{0z}=0.
\end{eqnarray}
Respectively, for $D^{K}$ we have:
\begin{eqnarray}\label{DD-K}
&&
D^K_2=\left(\pi_{\uparrow\downarrow}^K\zeta_{\uparrow\downarrow}+
\pi_{\downarrow\uparrow}^K\zeta_{\downarrow\uparrow}\right) F^2,\nonumber \\
&&
D^K_3=\left(\pi_{\uparrow\downarrow}^K\zeta_{\uparrow\downarrow}-
\pi_{\downarrow\uparrow}^K\zeta_{\downarrow\uparrow}\right) F^2,\nonumber \\
\nonumber \\
&& D_{zz}^K=\frac{F^2
(\pi^K_{\uparrow\uparrow}+\pi^K_{\downarrow\downarrow})}
{(1-2 F \pi^R)(1-2 F \pi^A)},\nonumber \\
&& D_{00}^K=\frac{(U^\rho)^2
(\pi^K_{\uparrow\uparrow}+\pi^K_{\downarrow\downarrow})} {(1-2
U^\rho \pi^R)(1-2 U^\rho \pi^A)} \rightarrow
\frac{\pi^K_{\uparrow\uparrow}+\pi^K_{\downarrow\downarrow}}
{4\pi^R\pi^A}
,\nonumber \\
&& D_{z0}^K=\frac{U^\rho
F(\pi^K_{\uparrow\uparrow}-\pi^K_{\downarrow\downarrow})} {(1-2 F
\pi^R)(1-2 U^\rho \pi^A)} \rightarrow
-\frac{F(\pi^K_{\uparrow\uparrow}-\pi^K_{\downarrow\downarrow})}
{2\pi^A(1-2 F \pi^R)},\nonumber \\
&& D_{0z}^K=\frac{U^\rho F
(\pi^K_{\uparrow\uparrow}-\pi^K_{\downarrow\downarrow})} {(1-2
U^\rho \pi^R)(1-2 F \pi^A)} \rightarrow -\frac{F
(\pi^K_{\uparrow\uparrow}-\pi^K_{\downarrow\downarrow})} {2\pi^R
(1-2 F \pi^A)},\nonumber \\ \nonumber \\
&&\qquad\zeta_{\alpha\beta}=\frac{1}{(1-2 F \pi^A_{\alpha\beta})
(1-2 F \pi^R_{\alpha\beta})}.
\end{eqnarray}
In Eqs.~(\ref{DD-RA}),(\ref{DD-K}) we took the limit
$U^{\rho}\rightarrow\infty$ that corresponds to $|V({\bf
q})\pi^{R(A)}(\omega,{\bf q})|\gg 1$ which is always the case at
small enough $|{\bf q}|$, as $V({\bf q}\rightarrow 0)$ diverges
because of the long-range character of Coulomb interaction.

Substituting Eqs.~(\ref{sigma})-(\ref{DD-K}) into
Eq.~(\ref{keld}), neglecting interaction correction to the Green's
function ${\cal G}_{E}\approx G_{E}$ and using Eq.~(\ref{av}) for
the disorder average of still not averaged pair of Green's
functions we arrive at the main quantitative result of this paper
given by Eq.~(\ref{comb1})-(\ref{K3}).

{\bf 6. Conclusion.} We have shown above that the relaxation of
the spin-dependent electron energy distribution at the total spin
magnetization ${\cal M}=0$ differs qualitatively from the case
${\cal M}\neq 0$. It is only in this case that the complete
relaxation to a spin-independent Fermi distribution is possible
due to electron-electron interaction alone. And it is in this case
that the infrared catastrophe is encountered in the collision
integral for a quasi-1d disordered wire. As a result, an
anomalously fast relaxation to a spin-independent non-equilibrium
distribution happens well before the complete equilibrium is
reached. The corresponding collision integral responsible for such
a fast relaxation can be approximated as
\begin{equation}
\label{approxColl} {\cal K}_{\rm coll}\approx
-K_{\sigma}\,(h_{\sigma,E}-h_{-\sigma,E}),
\end{equation}
where
\begin{equation}
\label{K-sigma} K_{\sigma}=\frac{C(F)}{2\pi\nu
S}\;\sqrt{\frac{\tau_{0}}{2D}}\,\int
(1-h_{\sigma,E'}h_{-\sigma,E'})dE',
\end{equation}
with $C(F)=\frac{F^{2}}{(1+\sqrt{1+F})(2+F)\sqrt{1+F}}$.

A remarkable feature of Eq.~(\ref{K-sigma}) emerging due to the
infrared catastrophe  is the quasi-elastic form of the collision
integral. If the infrared cut-off $1/\tau_{0}$ in the divergent
integral $\int_{1/\tau_{0}} \frac{d\omega}{|\omega|^{3/2}}$ is
small compared to the effective temperature, one can set
$\omega=0$ in the distribution functions $f_{\sigma}$ entering
Eq.~(\ref{comb3}). Thus we arrive at Eq.~(\ref{approxColl}) which
dependence on $f_{\sigma}(E)$ is identical to the elastic part of
the collision integral due to magnetic impurities. The only
difference is that the coefficient $K_{\sigma}$ also depends on
the integral of the distribution functions. Thus the triplet part
of the electron-electron interaction acts in this case similar to
the magnetic impurities.


\begin{thebibliography}{99}
\bibitem{foot1}Here and below we assume the case of dirty metal where the
distribution function $f(E;x,t)$ depends only on energy $E$ and
(slowly) on the space and time coordinates $x,t$.
\end{thebibliography}
\end{document}